# Bend-Net: Bending Loss Regularized Multitask Learning Network for Nuclei Segmentation in Histopathology Images

Haotian Wang[1], Aleksandar Vakanski[1], Changfa Shi[2], Min Xian[1]

[1]Department of Computer Science, University of Idaho, Idaho Falls, ID 83402 USA

[2]Mobile E-business Collaborative Innovation Center of Hunan Province, Hunan University of Technology of Business, Changsha 410205, Hunan, China.

**Abstract**

Separating overlapped nuclei is a major challenge in histopathology image analysis. Recently published approaches have achieved promising overall performance on nuclei segmentation; however, their performance on separating overlapped nuclei is quite limited. To address the issue, we propose a novel multitask learning network with a bending loss regularizer to separate overlapped nuclei accurately. The newly proposed multitask learning architecture enhances the generalization by learning shared representation from three tasks: instance segmentation, nuclei distance map prediction, and overlapped nuclei distance map prediction. The proposed bending loss defines high penalties to concave contour points with large curvatures, and applies small penalties to convex contour points with small curvatures. Minimizing the bending loss avoids generating contours that encompass multiple nuclei. In addition, two new quantitative metrics, Aggregated Jaccard Index of overlapped nuclei (AJIO) and Accuracy of overlapped nuclei (ACCO), are designed for the evaluation of overlapped nuclei segmentation. We validate the proposed approach on the CoNSeP and MoNuSegv1 datasets using seven quantitative metrics: Aggregate Jaccard Index, Dice, Segmentation Quality, Recognition Quality, Panoptic Quality, AJIO, and ACCO. Extensive experiments demonstrate that the proposed Bend-Net outperforms eight state-of-the-art approaches.

*Keywords:* Histopathology images analysis, nuclei segmentation, bending loss, multitask deep learning, cancer diagnosis.

## 1. Introduction

Histopathology nuclei segmentation aims to extract all nuclei from histopathology images, and it provides reliable evidence for cancer evaluation. Conventionally, pathologists examine the shapes and distributions of the nuclei under microscopes to determine the carcinoma and the malignancy level (He et al., 2011). However, the large number of nuclei makes the whole process time-consuming, low-throughput, and prone to human error. Automated nuclei segmentation is highly desirable in clinical practice. Recently, with the growing interest of digital pathology, the whole slide scanner provides a solution that transfers glass slides digitally to whole slide images (Pantanowitz, 2010).

In an H&E-stained histopathology image, nuclei are the first and the most visible structures among tissues. Accurate nuclei segmentation is essential in the further quantitative analysis (Aeffner et al., 2019), e.g., movement tracking, morphological changing, and nuclei counting. Many computational approaches have been proposed for automatic nuclei segmentation in histopathology images. Some conventional approaches (Ali et al., 2012; Cheng et al., 2009; Yang et al., 2006) utilized thresholding and watershed algorithms to segment nuclei, but these approaches are not robust in handling images with various nucleus types, fat tissue, and staining procedures. In recent years, deep learning-based approaches have been thriving in numerous biomedical image processing tasks (Badrinarayanan et al., 2017; Ronneberger et al., 2015; Shelhamer et al., 2015), and have achieved promising results in nuclei segmentation (Chen et al., 2016; Graham et al., 2018; Graham et al., 2019; Koohbanani et al., 2019; Kumar et al., 2017; Naylor et al., 2019; Oda et al., 2018; Qu et al., 2019; Vu et al., 2019; Xing et al., 2015; Zeng et al., 2019; Zhou et al., 2019). Xing et al. (2015) proposed a convolution neural network (CNN) to produce probability maps, and improved the robustness by using postprocessing, e.g., distance transformation, H-minima thresholding and region growing algorithm. Kumar et al. (2017) demonstrated a three-class (instance, boundary, and background) CNN that computes the label for each pixel to segment the nuclei. Naylor et al. (2019) constructed a DIST-map that utilized the regression output of nuclei distance maps for accurate nuclei segmentation. Although these methods achieved better results compared to conventional approaches, however, it is still challenging to segment nuclei accurately due to the existence of a large amount of overlapped nuclei.

Overlapped nuclei segmentation is challenging because of the lack of clear boundaries among nuclei, similar background textures, and large size and morphology variations. In recent published deep learning-based approaches, three main strategies have been proposed to address this challenge. *The first strategy* utilized the neural network to split the overlapped nuclei by generating both nuclei regions and boundaries. For example, Kumar *et al.* (2017) proposed a three-classes CNN, including instances, boundaries, and background to segment the overlapped nuclei. Chen *et al.* ( 2016) proposed a multitask learning framework that output instance map and boundary map in separate branches. Vu *et al*. (2019) constructed a multiscale deep residual network with instances and boundary classes to segment nuclei. *The second strategy* integrated features from overlapped nuclei to improve the overall segmentation performance. Zhou *et al.* ( 2019) proposed the CIA-Net that utilized spatial and texture dependencies between nuclei and contours to improve the robustness of nuclei segmentation. Koohbanahi *et al.* (2019) proposed a SpaNet that captures spatial features in a multiscale neural network. Graham *et al.* (2018) proposed a new weighted cross-entropy loss that was sensitive to the Hematoxylin stain. Qu *et al.* (2019) constructed a new cross-entropy loss to learn spatial features for improving localization accuracy. *The third strategy* utilized the Watershed algorithm to segment the overlapped nuclei. Naylor *et al.* (2019) constructed a regression network that generated markers for the Watershed algorithm to segment overlapped nuclei. Graham *et al*. (2019) proposed the HoVer-Net architecture to output the instance map and horizontal and vertical nuclei distance maps for obtaining the markers of the Watershed algorithm. According to the reported results, these approaches achieved better overall performance than conventional methods, but their ability to separate overlapped nuclei is still limited (Fig. 1).

To address the challenges above, we proposed a novel bending loss regularized deep multitask network for nuclei segmentation. *First*, the proposed multitask network consists of three decoder branches: 1) instance segmentation branch, 2) boundary-distance branch for all nuclei, and 3) boundary-distance branch for overlapped nuclei. The third branch is designed to identify overlapped nuclei. *Second*, we propose the bending energy-based regularizer to penalize large curvatures of nuclei contours. In histopathology images, the curvatures of nucleus contour points change smoothly; but if one contour contains two or multiple overlapped or touching nuclei, their touching points on the contour will have sharp curvature changes (Fig. 2). Inspired by this observation, we develop the bending loss to generate large penalties for contour points with large curvatures. *Third*, we propose two new metrics to evaluate overlapped nuclei segmentation. Previous approaches evaluate overlapped nuclei segmentation using metrics for overall segmentation performance, which hides the real performance of the overlapped nuclei segmentation. Compared to the closest work, HoVer-Net (Graham et al., 2019), both the proposed approach and the HoVer-Net follow the multitask learning architecture and use ResNet-50 as the building blocks. There are two major differences between the two approaches: 1) the proposed method introduced a new decoder branch to give focuses on overlapped nuclei; and 2) we propose the bending loss to penalize large curvatures of nuclei contours.

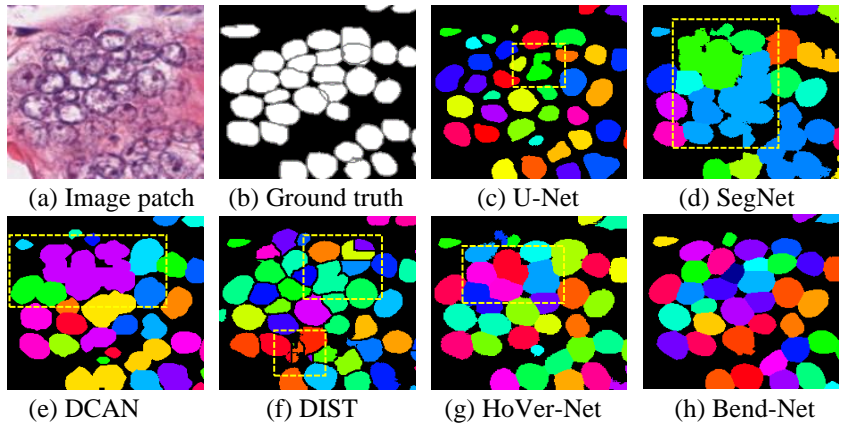

Fig. 1. Examples of state-of-the-art approaches in segmenting overlapped nuclei.

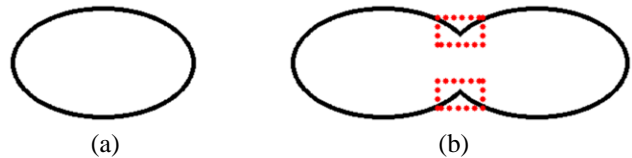

Fig. 2. Two contours. (a) An ideal nucleus contour; and (b) a contour contains two nuclei. Red rectangles highlight the touching points on the contour.

The rest of the paper is organized as follows. Section 2 describes the proposed method, including the bending loss, multitask learning network, and the loss function of the proposed architecture. Section 3 firstly describes the datasets and evaluation metrics in our experiments, then presents the implementation and training process. The experimental results are discussed in Sections 3.3 to 3.7. Section 4 is the conclusion and future work.

## 2. Method

The proposed method, namely Bend-Net, consists of two key components: the bending loss and multitask learning architecture. Firstly, we propose a bending energy-based regularizer for penalizing touching nuclei points. Secondly, we propose a multitask learning network with three decoder branches that focus on overlapped nuclei contours. The final loss function consists of the regular segmentation loss (Graham et al., 2019), overlapped nuclei loss, and the bending loss.

### 2.1 Bending loss

Bending energy has been widely applied in measuring the shapes of biological structures, e.g., blood cells (Canham et al., 1970), cardiac (Duncan et al., 1991), vesicle membranes (Du et al., 2006), and blood vessels (Stuhmer et al., 2013). Young *et al.* (1974) used the chain-code representations to model bending energy. Vliet *et al*. (1993) used the derivative-of-Gaussian filter to model bending energy in the gray-scale image for motion tracking. Wardetzky *et al.* (2008) modeled the discrete curvature and bending loss both in kinematic and dynamical treatment to solve the smoothness problem.

For 2D digital images, a contour is composed of discrete pixels, and the curvature of a contour point is computed by using the vectors created by neighboring points on the contour. For histopathology images, a nucleus usually has a smooth contour, and the points on the contour have small curvature changes; the points on the contour with large curvature have high probability to be the touching points of two/multiple nuclei (Fig. 2.). To split the touching nuclei, we define the bending loss that gives high penalties to the contour points with large curvatures, and small penalties to points with small curvatures. The proposed total loss is given by

$$L = L_0 + \alpha \cdot L_{be} \quad (1)$$

where $L_0$ refers to conventional segmentation loss (Section 2.3); $L_{be}$ denotes the proposed bending loss; and the parameter $\alpha$ controls the contribution of the bending loss. Let $C = \{c_i\}_{i=1}^{m}$ be the set of contour points of nuclei in an image, and $L_{be}$ is defined by

$$L_{be}(C) = \frac{1}{m}\sum_{i=1}^{m} BE(i) \qquad (2)$$

where $BE(i)$ is the discrete bending energy at the point $c_i$,

$$BE(i) = \frac{\kappa(i)^2 \left((1-\delta(c_i)) + \delta(c_i)\cdot \mu\right)}{|v(i,i+1)| + |v(i-1,i)|}, \qquad (3)$$

$$\kappa(i) = \frac{2|v(i-1,i) \times v(i,i+1)|}{|v(i-1,i)||v(i-1,i)| + v(i-1,i)\cdot v(i,i+1)} \qquad (4)$$

In Eq. (3), $\delta(c_i)$ is 1 if $c_i$ is a concave contour point, and 0 if $c_i$ is a convex point; $\kappa(i)$ is the curvature at $c_i$. For three consecutive pixels on a nucleus boundary with coordinates $x_{i-1}$, $x_i$ and $x_{i+1}$, $v(i-1, i)$ is the edge vector from point $i-1$ to $i$, such that $v(i-1, i) = x_i - x_{i-1}$; and $v(i, i+1)$ is the edge vector from $i$ to $i+1$, such that $v(i, i+1) = x_{i+1} - x_i$. Operator $|\cdot|$ calculates the length of a vector. $\mu$ defines the weight for concave contour points.

The 8-neighborhood system is applied to search neighbors for contour points. Ideally, a contour point only has two neighboring points, and their coordinates are used to calculate the edge vectors in Eqs. (3) and (4). As shown in Fig. 3, a point with eight neighbors has 28 combinations of possible curve patterns. All curve patterns are divided into five groups; in each group, the concave points and the convex points have different discrete bending loss values. In the first group, the four patterns construct straight-line segments, and their bending losses are all 0s. The second group shows patterns with $3\pi/4$ angle between edge vectors, and their bending losses are relatively small. In the last group, the eight patterns have large curvatures, and their bending losses are the largest in all patterns. The third and fourth groups illustrate patterns with the same angles between edge vectors, but they have different bending loss due to the different vector lengths.

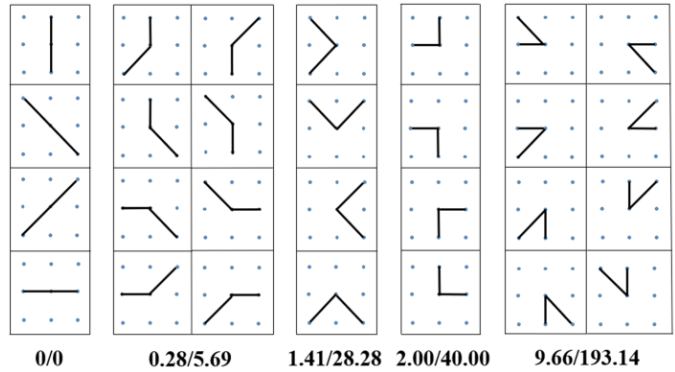

Fig. 3. Discrete bending losses for different curve patterns. In the value pairs 'A/B', 'A' represents the convex bending loss for the center point, and 'B' denotes the concave bending loss. The value rounds to two decimal places.

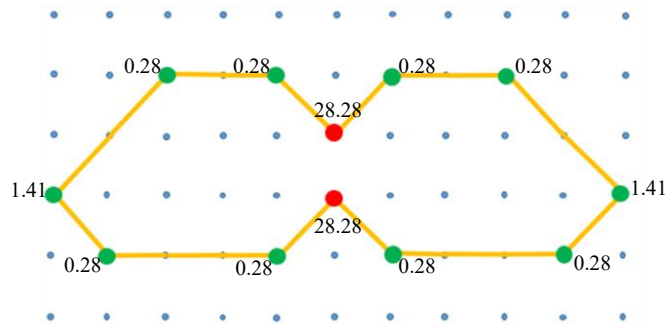

Fig. 4. A contour with both concave and convex points. Red dots highlight the concave points, and green dots highlight the convex points.

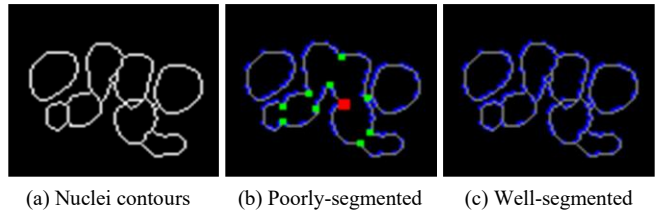

(a) Nuclei contours    (b) Poorly-segmented    (c) Well-segmented

Fig. 5. Different bending losses of different segmentation results. (a) Ground truth of eight nuclei contours; (b) bending losses of contour points of poorly-segmented nuclei; and (c) bending losses of well-segmented nuclei. Red: BE = 193.14, green: BE=28.28 and BE=40.0, blue: BE≤9.66, and grey: BE=0.

To determine the concave and convex points, the mid-point of two extended neighboring points is calculated. If the mid-point is out of the predicted nucleus, we define it as a concave point; otherwise, the point is a convex point. The concave points are more likely to be overlapped contour points, and the convex points are usually regular/normal points. Eq. (3) gives a larger penalty to concave points. The previous approach (Wang et al, 2020) calculated bending loss using curvature directly. Points with the same curvatures could be convex or concave; and convex points are more likely regular contour points, and concave points are likely to be overlapped contour points. The previous approach cannot distinguish convex and concave contour points and tends to over-segment nuclei.

A sample of overlapped nucleus contour is shown in Fig. 4. The red dots highlight the concave points and the green dots highlight the convex points. The concave points' bending loss values are 28.28. The mid-points of green dots are inside of the predicted nucleus, and they are convex points; their bending loss values are less than 1.41. In Fig. 4, the concave points with bending loss 28.28 and the convex points with bending loss 1.41 have the same curve pattern; however, the concave points produce 20 times as much loss as the convex points.

The proposed bending loss is rotation invariant since all patterns with the same angle between two edge vectors have the same bending loss. In practice, if two nuclei contours share some contour segments, one contour point may have more than two neighbors. In this scenario, we calculate the bending loss for all possible combinations, and choose the smallest loss as the discrete bending loss for the point.

As shown in Fig. 5, for poorly-segmented nuclei contours, all these touching contour points have relatively high (red and green points) bending losses. If the touching nuclei are well separated (Fig. 5(c)), and the bending loss of all contour points are less than 9.66.

## 2.2 Multitask learning network

The proposed multitask learning architecture is shown in Fig. 6. The network follows an encoder-decoder design, and has three decoder branches. The encoder employs ResNet-50 (He et al., 2016) as a feature extractor. In the first convolutional layer, 64 7×7 kernels with a stride of 1 are applied, but the following max-pooling layer is removed to preserve more information. The network has three decoders/tasks. The first task predicts the nuclei instance map (INST); the second produces each nucleus's horizontal and vertical boundary-distance map (HV); and the third ouputs the overlapped nuclei's horizontal and vertical boundary-distance map (OHV). All decoders in the three branches have the same sub-architectures and the dense units (Gao et al., 2017). The OHV and HV branches share weights through skip connections.

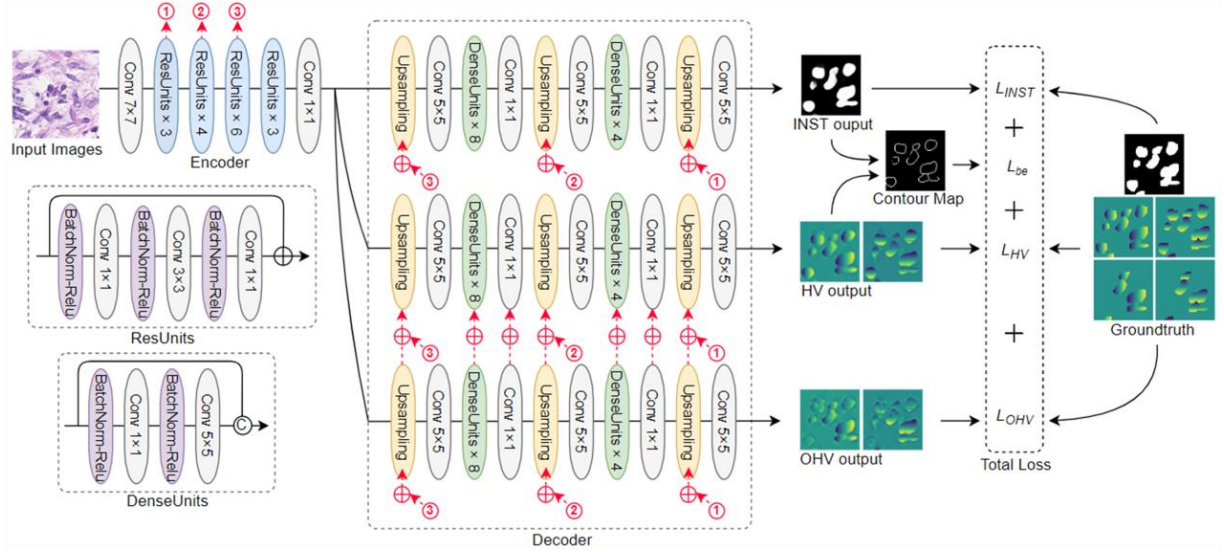

Fig. 6. Overview of the proposed Bend-Net. ⊕ denotes the summation; © denotes the concatenation; red arrow represents the skip connections; number with red circle denotes the connected position of skip-connections.

The weight-sharing among decoders are designed to use features learned from similar tasks. In traditional multitask learning networks, different branches typically addressed different tasks. However, in the proposed network, both the HV and OHV branches share some comment results; one for all nuclei, and the other for overlapped nuclei. To take advantage of the features from two similar tasks, we design the skip connections among two branches to share weights. Specifically, the network first learns the distance maps of overlapped nuclei and aggregates them through skip connections to distance maps in the HV branch.

**2.3 Loss function**

As shown in Fig. 6, the loss function of the proposed network has four terms: the losses from three different decoders and the proposed bending loss. Let $L_{INST}$ denote the loss of the binary instance map; $L_{HV}$ be the loss of the horizontal and vertical distance maps from the HV branch; and $L_{OHV}$ denote the loss of the horizontal and vertical distance maps from the OHV branch; $L_{be}$ is the bending loss. The proposed loss function also can split into the segmentation loss ($L_0$) and the bending loss regularizer (Eq. (1)). The total loss is given by:

$$L = \underbrace{L_{INST} + L_{HV} + L_{OHV}}_{L_0} + \alpha \cdot L_{be} \qquad (5)$$

where $\alpha$ is the weight of the bending loss in all loss function. We follow the design in (Graham et al., 2019) to set the losses of the three branches to have equal contributions to the total loss.

**Loss of the INST branch.** To segment the nuclei instance, we calculate the binary classification for each image pixel. $I$ and $I^*$ are the predicted instance map and the ground truth instance map for all nuclei. The loss ($L_{INST}$) is a summation of the cross-entropy loss ($L_{CE}$) and Dice loss ($L_{Dice}$). They are given by

$$L_{INST}(I, I^*) = L_{CE}(I, I^*) + L_{Dice}(I, I^*) \tag{6}$$

$$L_{CE}(I, I^*) = -\frac{1}{n}\sum_{i}^{n} I_i^* \log(I_i) \tag{7}$$

$$L_{Dice}(I, I^*) = 1 - \frac{2 \times \sum_{i}^{n} I_i I_i^*}{\sum_{i}^{n} I_i + \sum_{i}^{n} I_i^*} \tag{8}$$

where $I_i$ is the class prediction at point $i$, and $n$ denotes the number of pixels in an image patch. The INST branch separates the nuclei instance from the background.

**Loss of the HV branch.** The loss function is to compare the predicted distance maps ($D$) with the ground truth distance maps ($D^*$) for all nuclei. We employed the distance loss function in (Graham et al., 2019). The distance loss function is defined by

$$L_{dist}(D, D^*) = L_{Mse}(D, D^*) + 2 \cdot L_{Msge}(D, D^*) \tag{9}$$

$$L_{Mse}(D, D^*) = \frac{1}{n}\sum_{i}^{n} d_i^2 \tag{10}$$

$$L_{Msge}(D, D^*) = \frac{1}{n}\sum_{i}^{n} (\nabla_{d_i})^2 \tag{11}$$

where $L_{Mse}$ is the mean square error loss and $L_{Msge}$ is the mean square gradient error loss; $d$ is $D - D^*$, and $\nabla$ denotes the gradient calculation.

**Loss of the OHV branch.** $L_{OHV}$ is also defined using mean square error and the mean square gradient error (Eq. (9)). But $L_{OHV}$ is calculated using the predicted distance maps ($D$) and the ground truth distance maps of overlapped nuclei.

## 3. Experimental Results and Discussion

### 3.1 Datasets and evaluation metrics

In this paper, we validate the proposed method using two histopathology nuclei datasets: CoNSeP (Graham et al., 2019) and MoNuSegv1 (Kumar et al, 2017). CoNSeP is provided by the University of Warwick, and has 41 H&E-stained images from 16 colorectal adenocarcinoma (CRA) WSIs collected using Omnyx VL120 scanner. Six types of nuclei, normal epithelial, tumor epithelial, inflammatory, necrotic, muscle, and fibroblast exist in the dataset. The dataset contains 24,319 manually annotated nuclei (13,256 overlapped). The image size is $1000 \times 1000$ with the magnification at 40×. In the experiment, 27 images are utilized for training and validation, and 14 images for testing. The training and validation sets have 15,582 nuclei, and the test set has 8,791 nuclei.

MoNuSegv1 contains 30 images from TCGA (The Cancer Genomic Atlas) dataset. The original size of the images is $1000 \times 1000$, and there are more than 21,000 manually annotated nuclei from the breast, liver, kidney, prostate, bladder, colon, and stomach. The magnification is at 40×. In experiments, 16 images (4 breasts, 4 livers, 4 kidneys, 4 prostates) are used for training and validation, and 14 images for testing. The training and validation sets contain over 13,000 nuclei (4,431 overlapped), and the test set has 6,000 nuclei (2,436 overlapped). The author recently extended the dataset and published in (Kumar et al., 2020); however, it was not adopted in this study because it contains much less overlapped nuclei in their new test set compared with the previous test set (Kumar et al, 2017).

We employed five quantitative metrics to evaluate the overall performance of nuclei segmentation approaches: Aggregate Jaccard Index (AJI) (Kumar et al, 2017), Dice coefficient (Dice et al., 1945), Recognition Quality (RQ) (Alexander et al., 2019), Segmentation Quality (SQ) (Alexander et al., 2019), and Panoptic Quality (PQ) (Alexander et al., 2019). We propose two new metrics to evaluate the overlapped nuclei segmentation: Aggregated Jaccard Index of overlapped nuclei and accuracy for overlapped nuclei.

Let $G = \{G_i\}_{i=1}^{N}$ be the nuclei ground truth of an image, $N$ denote the total amount of segments in $G$; and let $S = \{S_k\}_{k=1}^{M}$ be the predicted segments of the corresponding image, $M$ denote the total amount of segments in $S$. AJI is an aggregate version of Jaccard Index and is defined by

$$AJI = \frac{\sum_{i=1}^{N} G_i \cap S_j}{\sum_{i=1}^{N} G_i \cup S_j + \sum_{S_k \in U} S_k}$$

where $S_j$ is the matched predicted segments that produce the largest Jaccard Index value with $G_i$; and $U$ denotes the set of unmatched predicted segments, where the total amount of $U$ is $(M - N)$.

Dice coefficient (DICE) is utilized to evaluate overall segmentation performance, the DICE is given by

$$DICE = \frac{2|G \cap S|}{(|G| + |S|)}$$

where operator $|\cdot|$ denotes the cardinalities of the segments.

PQ is used to estimate both detection and segmentation results. RQ is the familiar F1-score, and SQ is known as the average Jaccard Index of matched pairs. *RQ, SQ, PQ* are defined as

$$RQ = \frac{TP}{TP + \frac{1}{2}FP + \frac{1}{2}FN}$$

$$SQ = \frac{\sum_{(p,g) \in TP} IoU(p,g)}{TP}$$

$$PQ = RQ \times SQ$$

where $p$ refers to prediction, $g$ refers to the ground truth. The matched pairs $(p,g)$ are mathematically proven to be *unique matching* (33) if their IoU$(p,g)$ > 0.5. The *unique matching* splits the prediction and ground truth into three sets: the number of matched pairs (*TP*), the number of unmatched predictions (*FP*), and the number of unmatched ground truths (*FN*).

**Metrics for Overlapped Nuclei Segmentation**. We improved the Aggregated Jaccard Index (AJI) and accuracy metrics, and proposed two new metrics to evaluate overlapped nuclei segmentation, namely, AJI of overlapped nuclei (AJIO), accuracy for overlapped nuclei (ACCO). Because of the existence of many non-overlapped nuclei in images, traditional evaluation metrics cannot accurately validate the performance of overlapped nuclei segmentation. The proposed two metrics exclude all non-overlapped nuclei and focus on the evaluation of overlapped nuclei. Let $G = \{G_i\}_{i=1}^{N}$ be the overlapped nuclei in a ground truth image; and $S = \{S_k\}_{k=1}^{M}$ be the nuclei in the output image. AJIO is defined by

$$\text{AJIO} = \frac{\sum_{i=1}^{N} G_i \cap S_j}{\sum_{i=1}^{N} G_i \cup S_j}$$

where $S_j$ is the matched nucleus in $S$ that produces the largest Jaccard Index value with $G_i$.

Let $M$ be the number of matched nuclei pairs between the segmentation and ground truth, and $O$ denote the total number of overlapped nuclei in an image. For each overlapped nuclei, we iterate all the predicted segments, and count two nuclei matched if their Jaccard Index value is larger than a threshold $\tau$ (0.5). The ACCO is given by

$$\text{ACCO} = \frac{M}{O}$$

The two metrics are general and can be applied to other overlapped object segmentation.

### 3.2 Implementation and training

The proposed approach is trained by using an NVIDIA Titan Xp GPU. The encoder was pretrained on ImageNet; and we trained the decoder for 100 epochs to obtain the initial parameters for the decoder branches. The network was further fine-tuned for 100 epochs on the nuclei training set. The size of the final output images is 80×80 pixels, and these output images are merged to form images with the same size (1000×1000) as the original images. In experiments, the initial learning rate is $10^{-4}$ and is reduced to $10^{-5}$ after 50 epochs. The batch size is 8 for training the decoder and 2 for fine-tuning the network. Moreover, processing an image of size 1000×1000 with our architecture takes about one second.

The input dimensionality of the network is 270×270×3. We prepare the training, validation, and test sets by extracting patches from images with 270×270 pixels size. During the training stage, data augmentation strategies, i.e., rotation, Gaussian blur, and median blur, are utilized for generating more images. The ground truths of overlapped nuclei are two or multiple individual nuclei have connected-component labeling. An example histopathology image, the ground truth of all nuclei, and the overlapped nuclei are demonstrated in Fig. 7.

The proposed scheme comprises three stages: 1) preprocessing; 2) training of the proposed multitask learning network; and 3) postprocessing. The preprocessing performs color normalization (Vahadane et al., 2016) to reduce the impact of variations from the H&E staining and scanning processes. The postprocessing described in Graham et al. (2019) is employed in this study, which applies Sobel operators to the distance maps to generate initial contour map; then the difference of initial nuclei contour map and nuclei instance map is used to generate markers; and finally, the watershed algorithm is applied to generate nuclei regions.

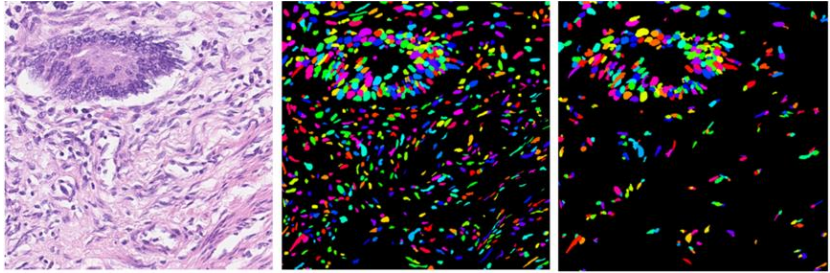

Fig. 7. Ground truth of an example image. From left to right: original image, ground truth of all nuclei, and ground truth of overlapped nuclei.

Table 1: Effectiveness of the proposed multitask learning architecture using the CoNSeP dataset.

| Methods | Metrics | | | | | |
| --- | --- | --- | --- | --- | --- | --- |
| | AJI | Dice | RQ | SQ | PQ | AJIO |
| Instance-Net | 0.371 | 0.841 | 0.603 | 0.771 | 0.471 | 0.296 |
| HoVer-Net | 0.545 | 0.840 | 0.674 | 0.773 | 0.522 | 0.520 |
| Ours-OHV* | 0.559 | 0.847 | 0.692 | 0.774 | 0.537 | 0.531 |
| Ours-skip* | **0.565** | **0.850** | **0.697** | **0.779** | **0.544** | **0.537** |

\* Ours-OHV denotes the proposed approach with the OHV branch; Our-skip has additional skip connections between the HV and OHV branches.

### 3.3. Effectiveness of the network architecture

The proposed multitask learning architecture uses HoVer-Net as the backbone and integrates our newly proposed overlapped nuclei (OHV) branch and skip connections (Fig. 6). To demonstrate the effectiveness of the proposed architecture, we compare the proposed network with the single-task network (Instance-Net), and two-task network (HoVer-Net). To perform a fair comparison, the proposed bending loss is not used. The approaches are evaluated on CoNSeP dataset by using AJI, Dice, RQ, SQ, and PQ scores. As shown in Table. 1, the Instance-Net does not apply any strategy to separate the overlapped nuclei and achieved very limited performance, e.g., AJI is only 0.371. The proposed network with the OHV branch ('Ours-OHV') achieved better average performance than the Instance-Net and HoVer-Net. With the new skip connections between the HV and OHV branches, the AJI, RQ, and PQ scores of the proposed approach

Table 2: Effectiveness of the proposed bending loss using the CoNSeP dataset.

| Methods | w/o bending loss | $L_{be}$ v1 | $L_{be}$ v2 | Metrics | | | | | |
|---|---|---|---|---|---|---|---|---|---|
| | | | | AJI | Dice | RQ | SQ | PQ | AJIO |
| HoVer-Net | ✓ | | | 0.545 | 0.840 | 0.674 | 0.773 | 0.522 | 0.520 |
| | | ✓ | | 0.552 | 0.844 | 0.683 | 0.774 | 0.530 | 0.523 |
| | | | ✓ | 0.559 | 0.846 | 0.690 | 0.776 | 0.537 | 0.528 |
| Ours | ✓ | | | 0.565 | 0.850 | 0.697 | 0.779 | 0.544 | 0.537 |
| | | ✓ | | 0.570 | 0.847 | 0.701 | 0.777 | 0.547 | 0.541 |
| | | | ✓ | **0.578** | **0.851** | **0.709** | **0.781** | **0.555** | **0.552** |

* $L_{be}$ v1 and $L_{be}$ v2 refer to our previous bending loss [35] and the newly proposed bending loss, respectively.

('Ours-skip') increased by 3.54%, 3.30%, and 4.04%, respectively. The AJIO scores demonstrate that the proposed approach outperforms HoVer-Net in separating overlapped nuclei.

### 3.4 Effectiveness of the Bending Loss

The newly proposed bending loss improves our original bending loss calculation in (Wang et al., 2020) by characterizing the difference between the concave and convex contour points. First, we compare the proposed multitask learning network without any bending loss, with the bending loss ($L_{be}$ v1) in (Wang et al., 2020), and with the newly proposed bending loss ($L_{be}$ v2). Second, we demonstrate the effectiveness of the bending loss by adding it to HoVer-Net. The CoNSeP dataset and AJI, Dice, RQ, SQ, and PQ scores are used in experiments. As shown in Table. 2, the proposed architecture with the *v1* bending loss (Wang et al., 2020) achieves better performance than that of the network without any bending loss, and the proposed architecture with the newly proposed v2 bending loss outperforms the network with the *v1* bending loss. The results demonstrate that the *v2* bending loss can improve the overall performance (AJI: from 0.565 to 0.578) of nuclei segmentation. Meanwhile, adding the *v1* or *v2* bending losses to HoVer-Net improves its overall performance, which demonstrated the potential of applying the bending loss to improve performance of other approaches. In addtion, the AJIO scores demonstrate that the proposed approach with the *v2* bending loss outperforms all other approaches in separating overlapped nuclei.

### 3.5 Parameter tuning

Two hyper-parameters, $\alpha\ and\ \mu$, exist in the proposed loss function. $\alpha$ balances the bending loss and all other losses (Eq. (5)); and $\mu$ (Eq. (3)) gives different weights to the concave and convex contour points when calculating the bending loss. We conducted grid search for the two parameters on the CoNSeP dataset by using the AJI score. Fig. 8 shows the AJI results of nine parameter combinations ($\mu$: 10, 20, 40; $\alpha$: 0.5, 1.0, 2.0). As shown in Fig. 8, the proposed approach achieved the best performance when $\mu$ is 20, and $\alpha$ is 1.0. Therefore, the bending loss of a concave curve pattern is twenty times the quantity of the same convex curve pattern. Refer to Fig. 3 for the bending loss of different curve patterns.

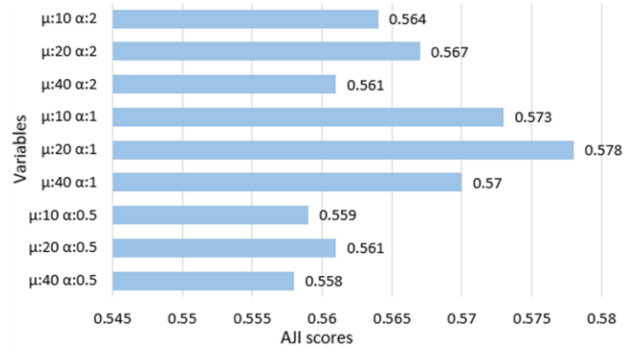

Fig. 8. Fine-tuning parameters using AJI scores.

Table 3: Overall test performance on the CoNSeP and MoNuSegv1 datasets.

| Methods | CoNSeP | | | | | MoNuSegv1 | | | | |
|---|---|---|---|---|---|---|---|---|---|---|
| | AJI | Dice | RQ | SQ | PQ | AJI | Dice | RQ | SQ | PQ |
| FCN8 | 0.289 | 0.782 | 0.426 | 0.697 | 0.297 | 0.426 | 0.779 | 0.592 | 0.708 | 0.421 |
| U-Net | 0.482 | 0.719 | 0.490 | 0.668 | 0.328 | 0.520 | 0.722 | 0.635 | 0.675 | 0.431 |
| SegNet | 0.461 | 0.699 | 0.482 | 0.667 | 0.322 | 0.508 | 0.797 | 0.672 | 0.742 | 0.500 |
| DCAN | 0.408 | 0.748 | 0.492 | 0.697 | 0.342 | 0.515 | 0.778 | 0.659 | 0.718 | 0.473 |
| DIST | 0.489 | 0.788 | 0.500 | 0.723 | 0.363 | 0.560 | 0.793 | 0.618 | 0.724 | 0.449 |
| Micro-Net | 0.531 | 0.784 | 0.613 | 0.751 | 0.461 | 0.581 | 0.785 | 0.700 | 0.737 | 0.517 |
| HoVer-Net | 0.545 | 0.840 | 0.674 | 0.773 | 0.522 | 0.606 | 0.818 | 0.765 | 0.767 | 0.588 |
| BEND | 0.553 | 0.846 | 0.683 | 0.776 | 0.530 | 0.627 | 0.827 | 0.770 | 0.766 | 0.590 |
| Bend-Net | **0.578** | **0.851** | **0.709** | **0.781** | **0.555** | **0.635** | **0.832** | **0.780** | **0.771** | **0.601** |

## 3.6 Performance comparison of state-of-the-art approaches

We compared eight deep learning-based approaches, including three widely used biomedical segmentation architectures: FCN8 (Shelhamer et al., 2015), U-Net (Ronneberger et al., 2015), and SegNet (Badrinarayanan et al., 2017), and five state-of-the-art nuclei segmentation approaches: DCAN (Chen et al., 2016), DIST (Naylor et al., 2019), Micro-Net (Raza et al., 2019), HoVer-net (Graham et al., 2019), and BEND (Wang et al. 2020). Table 3 shows the overall performance of nine approaches on two public datasets (CoNSeP and MoNuSegv1) using five metrics AJI, Dice, RQ, SQ, and PQ. Note that all other approaches are tested using the described experiment settings, and therefore, the values in Table 3 may not be the same as those reported in the original publications. The Watershed algorithm is applied to FCN8, U-Net, and SegNet for postprocessing, whereas the rest of the approaches are implemented by following the same strategy as in the original papers. As shown in Table III, the proposed method outperforms other eight approaches in terms of all five metrics. Among three general biomedical segmentation architectures, U-Net achieved the highest AJI and RQ scores, but it has lower Dice and SQ scores than those of FCN8. DCAN and DIST built upon FCN8 and U-Net, respectively. DCAN outperforms the FCN8 in all five metrics, and DIST outperforms U-Net in all five metrics. However, their overall segmentation performances still are limited. Micro-Net, HoVer-Net achieve comparative segmentation results. The proposed Bend-Net achieves better results than all other approaches on the two datasets in all five metrics.

Table 4: Overlapped nuclei segmentation performance on the CoNSeP and MoNuSegv1 Datasets.

| Methods | CoNSeP | | MoNuSegv1 | |
|---|---|---|---|---|
| | AJIO | ACCO | AJIO | ACCO |
| FCN8 | 0.350 | 0.328 | 0.337 | 0.358 |
| U-Net | 0.486 | 0.395 | 0.472 | 0.464 |
| SegNet | 0.411 | 0.262 | 0.407 | 0.406 |
| DCAN | 0.417 | 0.293 | 0.427 | 0.423 |
| DIST | 0.542 | 0.476 | 0.543 | 0.536 |
| Micro-Net | 0.513 | 0.495 | 0.513 | 0.504 |
| HoVer-Net | 0.520 | 0.558 | 0.542 | 0.613 |
| BEND | 0.529 | 0.561 | 0.553 | 0.627 |
| Bend-Net | **0.552** | **0.586** | **0.570** | **0.656** |

**3.7 Overlapped nuclei segmentation**

We proposed two new metrics, AJIO and ACCO, to evaluate overlapped nuclei segmentation. Table 4 shows the performance of overlapped nuclei segmentation on the CoNSeP and MoNuSegv1 datasets by using AJIO and ACCO scores. The DIST, Micro-Net, HoVer-Net, and the proposed method applied strategies to separate overlapped nuclei; therefore, their performances are significantly better than FCN8, U-Net, and SegNet. Our method achieved the best AJIO and ACCO scores on two datasets. Fig. 9 shows segmentation examples of six image regions with overlapped nuclei from the CoNSeP and MoNuSegv1 datasets. In the ground truth images, overlapped nuclei are represented using different colors. In the result images, if an approach can separate two overlapped nuclei, the two nuclei should be in two different colors. As shown in Fig. 9, FCN8, SegNet, and U-Net tend to miss small nuclei, and cannot separate overlapped nuclei. DCAN is slightly better in handling overlapped nuclei than FCN8, SegNet, and U-Net, but tends to miss nuclei that are not small. DIST separates overlapped nuclei better than the last four approaches, but it tends to over-segment nuclei and generate many small regions. Micro-Net performs well in segmenting overlapped nuclei; but it produces smaller nuclei regions than the ground truth and tends to over-segment nuclei. Hover-Net shows better segmentation results than other six approaches. It can segment out small nuclei, and the sizes of result nuclei regions are close to those of the ground truth. It outperforms other six approaches in segmenting overlapped nuclei; however, it has difficult in separating closely touched nuclei. In Fig. 9, the proposed method achieves the most accurate results on six images. The proposed approach not only can segment out small nuclei, but also can separate closely touched nuclei accurately.

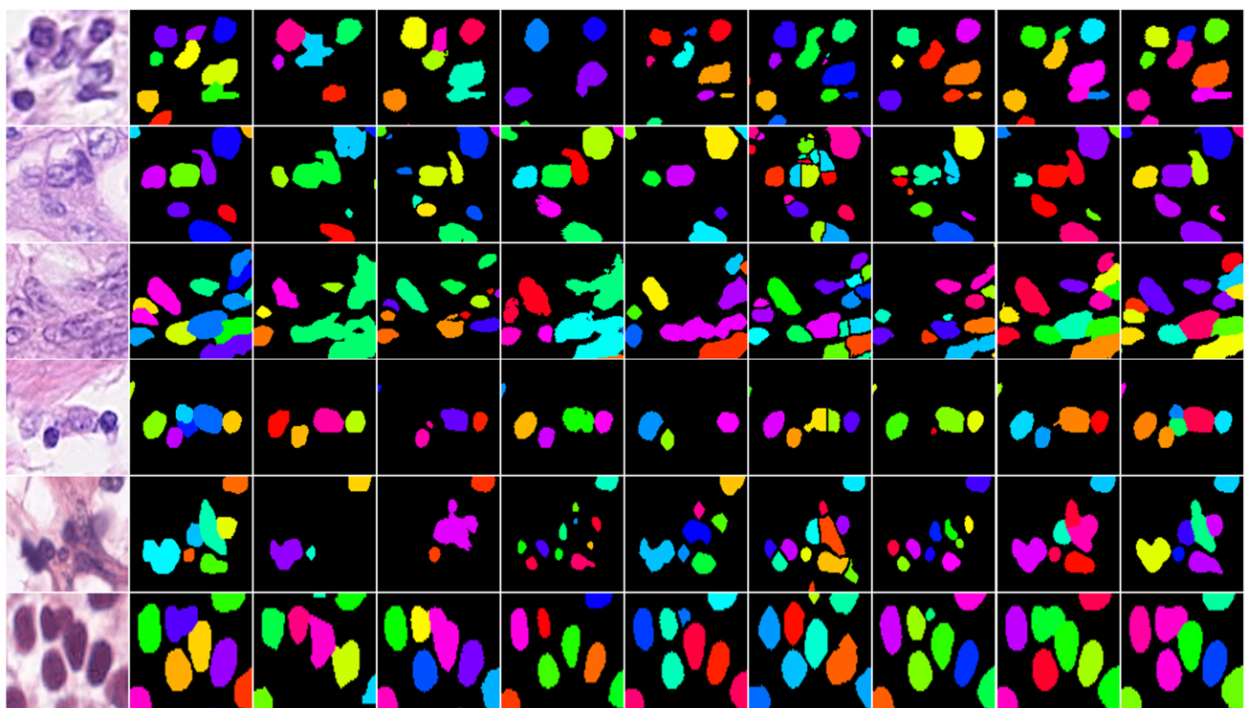

(a) Images    (b) Ground truth    (c) FCN8    (d) SegNet    (e) U-Ne    (f) DCAN    (g) DIST    (h) Micro-Net    (i) HoVer-Net    (j) Bend-Net
Fig. 9. Samples of comparative segmentation results for state-of-the-art models

## 4. Conclusion

In this paper, we propose a novel deep multitask learning network to address the challenge of segmenting overlapped nuclei in histopathology images. Firstly, we propose the bending loss regularizer, which defines different losses for the concave and convex points of nuclei contours. Experimental results demonstrated that the bending loss effectively improves the overall performance of nuclei segmentation, and it can also be integrated into other deep learning-based segmentation approaches. Secondly, the proposed multitask learning network integrates the OHV branch to learn knowledge from the overlapped nuclei, which enhances the segmentation of touching nuclei. Thirdly, we proposed two quantitative metrics, AJIO and ACCO, to evaluate overlapped nuclei segmentation. The extensive experimental results on two public datasets demonstrate that the proposed Bend-Net achieves state-of-the-art performance for nuclei segmentation. In the future, we will extend the proposed approach to more challenging tasks, such as gland segmentation and semantic image segmentation.

**Acknowledgments**

This work was supported, in part, by the Institute for Modeling Collaboration and Innovation (IMCI) at the University of Idaho through NIH Award #P20GM104420.

**References**

Aeffner, F., Zarella, M.D., Buchbinder, N., Bui, M.M., Goodman, M.R., Hartman, D.J., Lujan, G.M., Molani, M.A., Parwani, A.V., Lillard, K., 2019. Introduction to digital image analysis in whole-slide imaging: a white paper from the digital pathology association. Journal of pathology informatics 10.

Ali, S., Madabhushi, A., 2012. An integrated region-, boundary-, shape-based active contour for multiple object overlap resolution in histological imagery. IEEE transactions on medical imaging 31, 1448-1460.

Badrinarayanan, V., Kendall, A., Cipolla, R., 2017. Segnet: A deep convolutional encoder-decoder architecture for image segmentation. IEEE transactions on pattern analysis and machine intelligence 39, 2481-2495.

Bergou, M., Wardetzky, M., Robinson, S., Audoly, B., Grinspun, E., 2008. Discrete elastic rods, ACM SIGGRAPH 2008 papers, pp. 1-12.

Canham, P.B., 1970. The minimum energy of bending as a possible explanation of the biconcave shape of the human red blood cell. Journal of theoretical biology 26, 61-81.

Chen, H., Qi, X., Yu, L., Heng, P.-A., 2016. DCAN: deep contour-aware networks for accurate gland segmentation, Proceedings of the IEEE conference on Computer Vision and Pattern Recognition, pp. 2487-2496.

Cheng, J., Rajapakse, J.C., 2008. Segmentation of clustered nuclei with shape markers and marking function. IEEE Transactions on Biomedical Engineering 56, 741-748.

Dice, L.R., 1945. Measures of the amount of ecologic association between species. Ecology 26, 297-302.

Du, Q., Liu, C., Wang, X., 2006. Simulating the deformation of vesicle membranes under elastic bending energy in three dimensions. Journal of computational physics 212, 757-777.

Duncan, J.S., Lee, F.A., Smeulders, A.W., Zaret, B.L., 1991. A bending energy model for measurement of cardiac shape deformity. IEEE Transactions on medical imaging 10, 307-320.

Graham, S., Rajpoot, N.M., 2018. SAMS-NET: Stain-aware multi-scale network for instance-based nuclei segmentation in histology images, 2018 IEEE 15th International Symposium on Biomedical Imaging (ISBI 2018). IEEE, pp. 590-594.

Graham, S., Vu, Q.D., Raza, S.E.A., Azam, A., Tsang, Y.W., Kwak, J.T., Rajpoot, N., 2019. Hover-net: Simultaneous segmentation and classification of nuclei in multi-tissue histology images. Medical Image Analysis 58, 101563.

He, K., Zhang, X., Ren, S., Sun, J., 2016. Deep residual learning for image recognition, Proceedings of the IEEE conference on computer vision and pattern recognition, pp. 770-778.

He, L., Long, L.R., Antani, S., Thoma, G.R., 2012. Histology image analysis for carcinoma detection and grading. Computer methods and programs in biomedicine 107, 538-556.

Huang, G., Liu, Z., Van Der Maaten, L., Weinberger, K.Q., 2017. Densely connected convolutional networks, Proceedings of the IEEE conference on computer vision and pattern recognition, pp. 4700-4708.

Kirillov, A., He, K., Girshick, R., Rother, C., Dollár, P., 2019. Panoptic segmentation, Proceedings of the IEEE/CVF Conference on Computer Vision and Pattern Recognition, pp. 9404-9413.

Koohbanani, N.A., Jahanifar, M., Gooya, A., Rajpoot, N., 2019. Nuclear instance segmentation using a proposal-free spatially aware deep learning framework, International Conference on Medical Image Computing and Computer-Assisted Intervention. Springer, pp. 622-630.

Kumar, N., Verma, R., Anand, D., Zhou, Y., Onder, O.F., Tsougenis, E., Chen, H., Heng, P.-A., Li, J., Hu, Z., 2019. A multi-organ nucleus segmentation challenge. IEEE transactions on medical imaging 39, 1380-1391.

Kumar, N., Verma, R., Sharma, S., Bhargava, S., Vahadane, A., Sethi, A., 2017. A dataset and a technique for generalized nuclear segmentation for computational pathology. IEEE transactions on medical imaging 36, 1550-1560.

Long, J., Shelhamer, E., Darrell, T., 2015. Fully convolutional networks for semantic segmentation, Proceedings of the IEEE conference on computer vision and pattern recognition, pp. 3431-3440.

Naylor, P., Laé, M., Reyal, F., Walter, T., 2018. Segmentation of nuclei in histopathology images by deep regression of the distance map. IEEE transactions on medical imaging 38, 448-459.

Oda, H., Roth, H.R., Chiba, K., Sokolić, J., Kitasaka, T., Oda, M., Hinoki, A., Uchida, H., Schnabel, J.A., Mori, K., 2018. BESNet: boundary-enhanced segmentation of cells in histopathological images, International Conference on Medical Image Computing and Computer-Assisted Intervention. Springer, pp. 228-236.

Pantanowitz, L., 2010. Digital images and the future of digital pathology. Journal of pathology informatics 1.

Qu, H., Yan, Z., Riedlinger, G.M., De, S., Metaxas, D.N., 2019. Improving nuclei/gland instance segmentation in histopathology images by full resolution neural network and spatial constrained loss, International Conference on Medical Image Computing and Computer-Assisted Intervention. Springer, pp. 378-386.

Raza, S.E.A., Cheung, L., Shaban, M., Graham, S., Epstein, D., Pelengaris, S., Khan, M., Rajpoot, N.M., 2019. Micro-Net: A unified model for segmentation of various objects in microscopy images. Medical image analysis 52, 160-173.

Ronneberger, O., Fischer, P., Brox, T., 2015. U-net: Convolutional networks for biomedical image segmentation, International Conference on Medical image computing and computer-assisted intervention. Springer, pp. 234-241.

Stuhmer, J., Schroder, P., Cremers, D., 2013. Tree shape priors with connectivity constraints using convex relaxation on general graphs, Proceedings of the IEEE International conference on Computer Vision, pp. 2336-2343.

Vahadane, A., Peng, T., Sethi, A., Albarqouni, S., Wang, L., Baust, M., Steiger, K., Schlitter, A.M., Esposito, I., Navab, N., 2016. Structure-preserving color normalization and sparse stain separation for histological images. IEEE transactions on medical imaging 35, 1962-1971.

Verbeek, P., Van Vliet, L., 1993. Curvature and bending energy in digitized 2D and 3D images, 8th Scandinavian Conference on Image Analysis, Tromso, Norway.

Vu, Q.D., Graham, S., Kurc, T., To, M.N.N., Shaban, M., Qaiser, T., Koohbanani, N.A., Khurram, S.A., Kalpathy-Cramer, J., Zhao, T., 2019. Methods for segmentation and classification of digital microscopy tissue images. Frontiers in bioengineering and biotechnology 7, 53.

Wang, H., Xian, M., Vakanski, A., 2020. Bending loss regularized network for nuclei segmentation in histopathology images, 2020 IEEE 17th International Symposium on Biomedical Imaging (ISBI). IEEE, pp. 1-5.

Xing, F., Xie, Y., Yang, L., 2015. An automatic learning-based framework for robust nucleus segmentation. IEEE transactions on medical imaging 35, 550-566.

Yang, X., Li, H., Zhou, X., 2006. Nuclei segmentation using marker-controlled watershed, tracking using mean-shift, and Kalman filter in time-lapse microscopy. IEEE Transactions on Circuits and Systems I: Regular Papers 53, 2405-2414.

Young, I.T., Walker, J.E., Bowie, J.E., 1974. An analysis technique for biological shape. I. Information and control 25, 357-370.

Zeng, Z., Xie, W., Zhang, Y., Lu, Y., 2019. RIC-Unet: An improved neural network based on Unet for nuclei segmentation in histology images. Ieee Access 7, 21420-21428.

Zhou, Y., Onder, O.F., Dou, Q., Tsougenis, E., Chen, H., Heng, P.-A., 2019. Cia-net: Robust nuclei instance segmentation with contour-aware information aggregation, International Conference on Information Processing in Medical Imaging. Springer, pp. 682-693.